\documentclass[aps,prl,twocolumn,showpacs,groupaddress]{revtex4}
\usepackage{graphicx}% Include figure files
\usepackage{dcolumn}% Align table columns on decimal point
\usepackage{bm}% bold math
\usepackage{color}
\begin{document}
%\preprint{}
%%%%%%%%%%%%%%%%%%%%%%%%%%%%%%%%%%%%%%%%%%%%%%%%%%%%%%%%%%%%%%%%%%%%%%%%%%%%%%%%%%%%%%%%%%%%%%%%%%%
%
%%%%%%%%%%%%%%%%%%%%%%%%%%%%%%%%%%%%%%%%%%%%%%%%%%%%%%%%%%%%%%%%%%%%%%%%%%%%%%%%%%%%%%%%%%%%%%%%%%%
%REVISED INFORMATION
%2008/4/8/ replace 1c.eps with 1d.eps
\title{Electrons doped in cubic perovskite SrMnO$_{3}$: isotropic metal versus chainlike ordering of Jahn-Teller polarons}
%%%%%%%%%%%%%%%%%%%%%%%%%%%%%%%%%%%%%%%%%%%%%%%%%%%%%%%%%%%%%%%%%%%%%%%%%%%%%%%%%%%%%%%%%%%%%%%%%%%
%
%%%%%%%%%%%%%%%%%%%%%%%%%%%%%%%%%%%%%%%%%%%%%%%%%%%%%%%%%%%%%%%%%%%%%%%%%%%%%%%%%%%%%%%%%%%%%%%%%%%
\author{H. Sakai$^1$, S. Ishiwata$^{1,2}$, D. Okuyama$^1$, A. Nakao$^3$, H. Nakao$^3$, Y. Murakami$^3$, Y. Taguchi$^1$, and Y. Tokura$^{1,2,4}$}
%\email[]{sakai.hide@riken.jp}
%\thanks{}
\affiliation{$^1$Cross-Correlated Materials Research Group (CMRG) and Correlated Electron Research Group (CERG), Advanced Science Institute (ASI), RIKEN, Wako 351-0198, Japan\\
$^2$Department of Applied Physics, University of Tokyo, Tokyo 113-8656, Japan\\
$^3$Condensed Matter Research Center and Photon Factory, Institute of Materials Structure Science, KEK, Tsukuba, 305-0801, Japan\\
$^4$Multiferroics Project, ERATO, Japan Science and Technology Agency (JST), Tokyo 113-8656, Japan}
%\date{}
%%%%%%%%%%%%%%%%%%%%%%%%%%%%%%%%%%%%%%%%%%%%%%%%%%%%%%%%%%%%%%%%%%%%%%%%%%%%%%%%%%%%%%%%%%%%%%%%%%%
%
%%%%%%%%%%%%%%%%%%%%%%%%%%%%%%%%%%%%%%%%%%%%%%%%%%%%%%%%%%%%%%%%%%%%%%%%%%%%%%%%%%%%%%%%%%%%%%%%%%%
\begin{abstract}
Single crystals of electron-doped SrMnO$_{3}$ with a cubic perovskite structure have been systematically investigated as the most canonical (orbital-degenerate) double-exchange system, whose ground states have been still theoretically controversial.
With only 1-2\% electron doping by Ce substitution for Sr, a G-type antiferromagnetic metal with a tiny spin canting in a cubic lattice shows up as the ground state, where the Jahn-Teller polarons with heavy mass are likely to form.
Further electron doping above 4\%, however, replaces this isotropic metal with an insulator with tetragonal lattice distortion, accompanied by a quasi-one-dimensional $3z^{2}\!-\!r^{2}$ orbital ordering with the C-type antiferromagnetism.
The self-organization of such dilute polarons may reflect the critical role of the cooperative Jahn-Teller effect that is most effective in the originally cubic system.
\end{abstract}
%\keywords{}
%%%%%%%%%%%%%%%%%%%%%%%%%%%%%%%%%%%%%%%%%%%%%%%%%%%%%%%%%%%%%%%%%%%%%%%%%%%%%%%%%%%%%%%%%%%%%%%%%%%
%
%%%%%%%%%%%%%%%%%%%%%%%%%%%%%%%%%%%%%%%%%%%%%%%%%%%%%%%%%%%%%%%%%%%%%%%%%%%%%%%%%%%%%%%%%%%%%%%%%%%
\maketitle
%%%%%%%%%%%%%%%%%%%%%%%%%%%%%%%%%%%%%%%%%%%%%%%%%%%%%%%%%%%%%%%%%%%%%%%%%%%%%%%%%%%%%%%%%%%%%%%%%%%
%
%%%%%%%%%%%%%%%%%%%%%%%%%%%%%%%%%%%%%%%%%%%%%%%%%%%%%%%%%%%%%%%%%%%%%%%%%%%%%%%%%%%%%%%%%%%%%%%%%%%
Charge carriers doped in magnetic insulators have long attracted significant interest because of the possible application to spintronics \cite{Wolf2001Science} as well as of their underlying fundamental physics \cite{Imada1998RMP}.
There, the exchange coupling between conduction electrons and localized spins plays a crucial role, as represented by Kondo, RKKY, and double-exchange (DE) interactions.
Among them, the DE idea was first proposed by Zener \cite{Zener1951PRa} to explain the ferromagnetic (FM) interaction in hole-doped perovskite manganites and has been renewed throughout the intensive studies of colossal magnetoresistance phenomena \cite{Tokura2006review}.
A pioneering work by de Gennes \cite{Gennes1960PRa} suggested that the competition between the antiferromagnetic (AFM) superexchange and the FM DE results in the canted AFM ground state when carriers are doped.
Some of recent studies, however, argued that the phase-separated (FM-AFM-coexisting) state is stabilized rather than the homogeneous one \cite{Yunoki1998PRLa,Kagan1999EPJBa}.
It was further pointed out that the orbital degeneracy of $e_{g}$ bands favors the anisotropic AFM state such as a chainlike C-type structure \cite{Brink1999PRLa}.
Thus, the ground state of the orbital-degenerate DE model has been a longstanding problem to be clarified experimentally.
%%%%%%%%%%%%%%%%%%%%%%%%%%%%%%%%%%%%%%%%%%%%%%%%%%%%%%%%%%%%%%%%%%%%%%%%%%%%%%%%%%%%%%%%%%%%%%%%%%%
%
%%%%%%%%%%%%%%%%%%%%%%%%%%%%%%%%%%%%%%%%%%%%%%%%%%%%%%%%%%%%%%%%%%%%%%%%%%%%%%%%%%%%%%%%%%%%%%%%%%%
\par
\begin{figure}
\includegraphics[width=8.5cm]{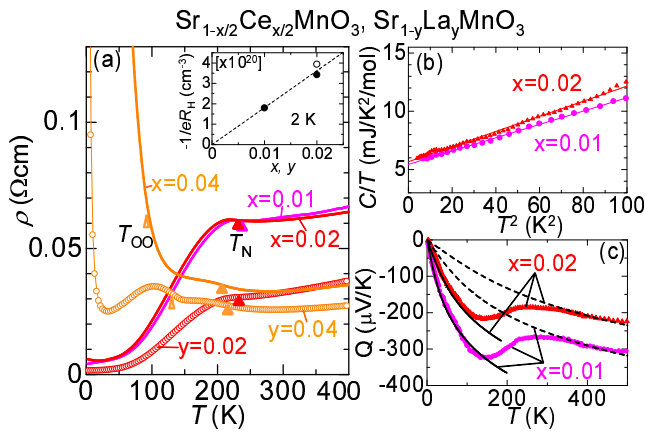}
\caption{\label{fig:polaron}(Color online) (a) $T$ profiles of $\rho$ at 0 T for single crystals of Sr$_{1-x/2}$Ce$_{x/2}$MnO$_{3}$ ($x$=0.01-0.04, solid lines) and Sr$_{1-y}$La$_{y}$MnO$_{3}$ ($y$=0.02-0.04, open circles). The closed and open triangles indicate $T_{\rm N}$(G) and $T_{\rm OO}$, respectively. Inset: effective carrier number at 2 K as a function of $x$ (closed circle) and $y$ (open circle) deduced from Hall coefficient $R_{\rm H}$. The dashed line corresponds to the nominal carrier density calculated from chemical composition. (b) Specific heat $C$ divided by $T$ is plotted against $T^{2}$ for $x$=0.01 and 0.02. (c) $T$ profiles of Seebeck coefficient $Q$. The solid and dashed lines in (c) indicate the calculated results using the Boltzmann equation \cite{Okuda2001PRBa}.}
\end{figure}

%%%%%%%%%%%%%%%%%%%%%%%%%%%%%%%%%%%%%%%%%%%%%%%%%%%%%%%%%%%%%%%%%%%%%%%%%%%%%%%%%%%%%%%%%%%%%%%%%%%
%
%%%%%%%%%%%%%%%%%%%%%%%%%%%%%%%%%%%%%%%%%%%%%%%%%%%%%%%%%%%%%%%%%%%%%%%%%%%%%%%%%%%%%%%%%%%%%%%%%%%
Most of the previous experimental studies on this issue have focused on electron-doped CaMnO$_{3}$ \cite{Martin1997JSSCa,Neumeier2000PRBa,Zeng2001PRBa,Caspi2004PRBa} with the GdFeO$_{3}$-type orthorhombic distortion that results in the partial lift of the $e_{g}$ orbital degeneracy.
SrMnO$_{3}$ with a cubic structure, on the other hand, forms the simplest and most ideal DE system.
Nevertheless, while SrMnO$_{3}$ in the high-electron-doping regime ($\sim$30-50\%) has been reported \cite{Kikuchi1999JSSCa,Sundaresan2000EPJBa,Mandal2004PRBa}, only few studies have been performed for this compound in the low-doping regime (below 5\%) \cite{Hervieu2000ChemMata,Chmaissem2003PRBa}.
One of the main reasons for this is the difficulty in synthesis of SrMnO$_{3}$ crystal with a cubic perovskite structure, since the hexagonal form is obtained under the conventional condition of solid-state reaction.
In this study, we have developed a method of synthesizing high-quality single crystals of cubic perovskite SrMnO$_{3}$ by combining a floating-zone method with high-pressure oxygen annealing.
Chemical substitution of Sr$^{2+}$ with Ce$^{4+}$ (La$^{3+}$) affords two (one) electron-type carriers \cite{Zeng2001PRBa}, and hence the carrier density for Sr$_{1-x/2}$Ce$_{x/2}$MnO$_{3}$ (Sr$_{1-y}$La$_{y}$MnO$_{3}$) corresponds to $x$ ($y$) per Mn site.
This was indeed confirmed by the Hall coefficient measurements [inset to Fig. \ref{fig:polaron}(a)].
By systematic transport, magnetic, and x-ray diffraction measurements on these crystals, we have experimentally unraveled the genuine phase diagram of the orbital-degenerate DE system.
As we show below, chainlike orbital ordering (OO) is realized in the ground state for SrMnO$_{3}$ with as small as 4\% electron doping (less than 1/3 for CaMnO$_{3}$), signifying the more stable insulating OO than in the CaMnO$_{3}$ system.
This behavior is totally opposite to the general trend observed in various manganites, where charge localization is more favored in a more distorted lattice.
These indicate a significant role of collective Jahn-Teller effects in the originally high-symmetry (cubic) system.
Below this doping concentration, furthermore, a G-type (staggered-type) AFM metal with a considerably renormalized mass manifests itself at the ground state, where strong electron-phonon coupling has been suggested by quantitative analyses based on a polaron model.
Thus, the itinerancy and self-organization of the dilute Jahn-Teller polarons critically compete in the lightly-doped regime.
%%%%%%%%%%%%%%%%%%%%%%%%%%%%%%%%%%%%%%%%%%%%%%%%%%%%%%%%%%%%%%%%%%%%%%%%%%%%%%%%%%%%%%%%%%%%%%%%%%%
%
%%%%%%%%%%%%%%%%%%%%%%%%%%%%%%%%%%%%%%%%%%%%%%%%%%%%%%%%%%%%%%%%%%%%%%%%%%%%%%%%%%%%%%%%%%%%%%%%%%%
\par
\begin{figure}
\includegraphics[width=6.5cm]{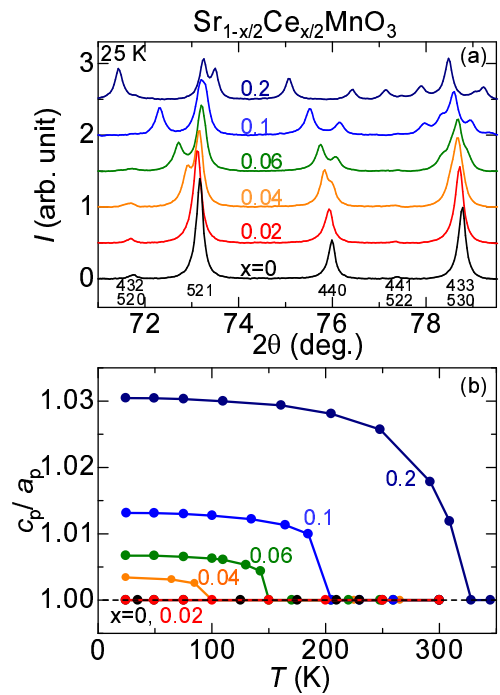}
\caption{\label{fig:x-ray}(Color online) (a) Synchrotron x-ray powder diffraction profiles for Sr$_{1-x/2}$Ce$_{x/2}$MnO$_{3}$ ($0\!\le\!x\!\le\!0.2$) at 25 K. The indices are based on the cubic setting ($Pm\bar{3}m$) and $a_{\rm p}$ monotonously increases with increasing $x$. (b) Temperature profiles of tetragonality ($c_{\rm p}/a_{\rm p}$) in warming runs.}
\end{figure}

%%%%%%%%%%%%%%%%%%%%%%%%%%%%%%%%%%%%%%%%%%%%%%%%%%%%%%%%%%%%%%%%%%%%%%%%%%%%%%%%%%%%%%%%%%%%%%%%%%%
%
%%%%%%%%%%%%%%%%%%%%%%%%%%%%%%%%%%%%%%%%%%%%%%%%%%%%%%%%%%%%%%%%%%%%%%%%%%%%%%%%%%%%%%%%%%%%%%%%%%%
Single crystals of SrMnO$_{3}$ were synthesized with the following two-step procedures.
We first synthesized single crystals of oxygen-deficient SrMnO$_{3-\delta}$ ($\delta\!\sim\!0.5$) with an orthorhombic structure by using a floating-zone method in an argon atmosphere.
Then, a piece of the single crystal ($\phi$3$\times$4 mm) sealed in a gold capsule with oxidizing agent (KClO$_{4}$) was treated at $\sim$6.5 GPa and $\sim$600$^{\circ}$C for 1h, using a conventional cubic anvil-type high-pressure apparatus.
%The pressure cell consists of a carbon heater and a pyrophyllite pressure transmitting medium.
The obtained  fully-oxidized SrMnO$_{3}$ remains single-crystalline and has a cubic structure [Fig. \ref{fig:x-ray}(a)].
For Ce- or La-doped compounds, the same synthesis process was applicable.
Synchrotron powder x-ray diffraction measurements with wavelength of 0.8260\AA were carried out at the Beam Line 8A at the Photon Factory, KEK, Tsukuba.
The magnetization ($M$) was measured with a superconducting quantum interference device (Quantum Design).
The four-probe resistivity ($\rho$), Hall coefficients ($R_{\rm H}$), specific heat ($C$), and Seebeck coefficients ($Q$) were measured using Physical Property Measurement System (Quantum Design).
The $Q$ measurement above 400 K was performed by AC modulating temperature gradient of 1-5 K.
%%%%%%%%%%%%%%%%%%%%%%%%%%%%%%%%%%%%%%%%%%%%%%%%%%%%%%%%%%%%%%%%%%%%%%%%%%%%%%%%%%%%%%%%%%%%%%%%%%%
%
%%%%%%%%%%%%%%%%%%%%%%%%%%%%%%%%%%%%%%%%%%%%%%%%%%%%%%%%%%%%%%%%%%%%%%%%%%%%%%%%%%%%%%%%%%%%%%%%%%%
\begin{figure}
\includegraphics[width=7cm]{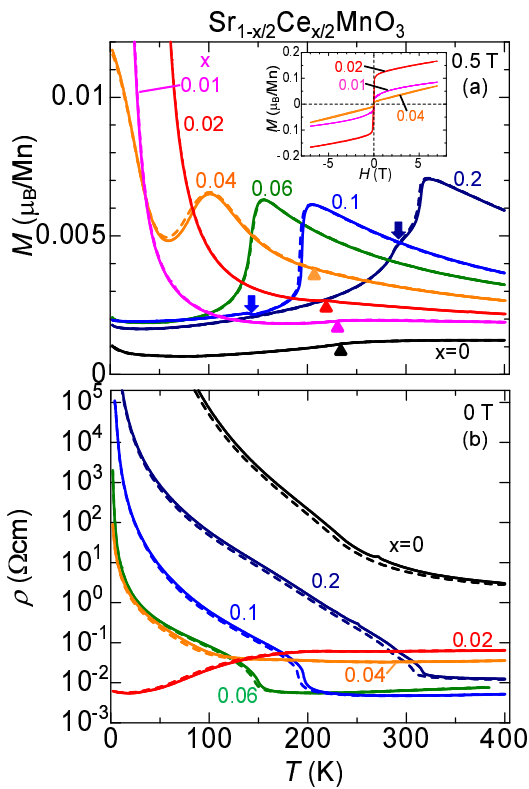}
\caption{\label{fig:MTRT}(Color online) Temperature ($T$) profiles of magnetization ($M$) at 0.5 T (a) and resistivity ($\rho$) at 0 T (b) for Sr$_{1-x/2}$Ce$_{x/2}$MnO$_{3}$ ($0\!\le\!x\!\le\!0.2$) single crystals. The solid and dashed lines correspond to warming and cooling runs, respectively. The closed triangles and arrows in (a) denote the N\'{e}el temperatures of G-type and of C-type antiferromagnetic (AFM) phases, respectively. Inset to (a): $M$ versus magnetic field ($H$) up to 7 T at 2 K for $x$=0.01-0.04.}
\end{figure}
%%%%%%%%%%%%%%%%%%%%%%%%%%%%%%%%%%%%%%%%%%%%%%%%%%%%%%%%%%%%%%%%%%%%%%%%%%%%%%%%%%%%%%%%%%%%%%%%%%%
%
%%%%%%%%%%%%%%%%%%%%%%%%%%%%%%%%%%%%%%%%%%%%%%%%%%%%%%%%%%%%%%%%%%%%%%%%%%%%%%%%%%%%%%%%%%%%%%%%%%%
\par
%%%%%%%%%%%%%%%%%%%%%%%%%%%%%%%%%%%%%%%%%%%%%%%%%%%%%%%%%%%%%%%%%%%%%%%%%%%%%%%%%%%%%%%%%%%%%%%%%%%
%
%%%%%%%%%%%%%%%%%%%%%%%%%%%%%%%%%%%%%%%%%%%%%%%%%%%%%%%%%%%%%%%%%%%%%%%%%%%%%%%%%%%%%%%%%%%%%%%%%%%
We show in Figs. \ref{fig:MTRT}(a) and (b) the overall temperature ($T$) profile of $M$ at 0.5 T and $\rho$ at 0 T for Sr$_{1-x/2}$Ce$_{x/2}$MnO$_{3}$ ($0\!\le\!x\!\le\!0.2$) single crystals, respectively.
The undoped SrMnO$_{3}$ is insulating over the whole $T$ region and shows a transition from paramagnetic (PM) phase to G-type AFM one at $T_{\rm N}$(G)$\sim$231 K [closed triangles in Fig. \ref{fig:MTRT}(a)].
Ce substitution for Sr by only 0.5-1\% makes the system metallic over the whole $T$ range [see also Fig. \ref{fig:polaron}(a)], while $T_{\rm N}$(G) remains at $\sim$220 K.
In these compounds, $M$ steeply increases toward the lowest $T$, which signifies the canting of the AFM spins.
In fact, the corresponding $M$-$H$ curves at 2 K [inset to Fig. \ref{fig:MTRT}(a)] exhibit the small spontaneous $M$: at most $\sim$0.12$\mu_{\rm B}$/Mn for $x$=0.02 corresponding to the canting angle $\sim$2.3$^{\circ}$.
Note here that the phase separation between the FM metal and AFM insulator can be ruled out, since the FM phase would amount to only 4\% in volume fraction, judging from the spontaneous $M$ value, which is well below the percolation threshold for metallic conduction \cite{phaseseparation}.
%%%%%%%%%%%%%%%%%%%%%%%%%%%%%%%%%%%%%%%%%%%%%%%%%%%%%%%%%%%%%%%%%%%%%%%%%%%%%%%%%%%%%%%%%%%%%%%%%%%
%
%%%%%%%%%%%%%%%%%%%%%%%%%%%%%%%%%%%%%%%%%%%%%%%%%%%%%%%%%%%%%%%%%%%%%%%%%%%%%%%%%%%%%%%%%%%%%%%%%%%
\par
%%%%%%%%%%%%%%%%%%%%%%%%%%%%%%%%%%%%%%%%%%%%%%%%%%%%%%%%%%%%%%%%%%%%%%%%%%%%%%%%%%%%%%%%%%%%%%%%%%%
%
%%%%%%%%%%%%%%%%%%%%%%%%%%%%%%%%%%%%%%%%%%%%%%%%%%%%%%%%%%%%%%%%%%%%%%%%%%%%%%%%%%%%%%%%%%%%%%%%%%%
With increasing $x$ up to 0.04, the metallic ground state is replaced by the insulating one.
For $x$=0.06-0.2, a distinct metal-insulator transition shows up, accompanied by a sudden jump in $\rho$ and a drop in $M$ as $T$ decreases.
The crystal structure also changes from cubic to tetragonal with the elongation of $c$ axis (Fig. \ref{fig:x-ray}), which indicates the $3z^{2}\!-\!r^{2}$-type OO.
Below this temperature, further anomalies were observed in the $M$-$T$ curves (arrows) for $x$$\ge$0.1, which suggests the C-type AFM transition \cite{Sundaresan2000EPJBa,Hervieu2000ChemMata,Chmaissem2003PRBa}.
The $3z^{2}\!-\!r^{2}$ electrons should be strongly confined within the one-dimensional chain also by the interchain AFM (C-type) order, and therefore the Coulombic correlation and/or electron-lattice interaction easily cause the self-trapping of the electrons as manifested by the highly insulating ground state.
The $x$=0.04 compound locates close to the metal-insulator phase boundary and shows reentrant spin and orbital transitions with decreasing $T$: PM$\rightarrow$G-type AFM$\rightarrow$OO (C-type AFM) [See also Fig. \ref{fig:polaron}(a)].
We further observed a rapid increase in $M$ below $\sim$40 K and small spontaneous $M$ ($\sim$0.01$\mu_{\rm B}$/Mn) at 2 K, which suggests the spin canting in the C-type AFM.
%%%%%%%%%%%%%%%%%%%%%%%%%%%%%%%%%%%%%%%%%%%%%%%%%%%%%%%%%%%%%%%%%%%%%%%%%%%%%%%%%%%%%%%%%%%%%%%%%%%
%
%%%%%%%%%%%%%%%%%%%%%%%%%%%%%%%%%%%%%%%%%%%%%%%%%%%%%%%%%%%%%%%%%%%%%%%%%%%%%%%%%%%%%%%%%%%%%%%%%%%
\begin{figure}
\includegraphics[width=8cm]{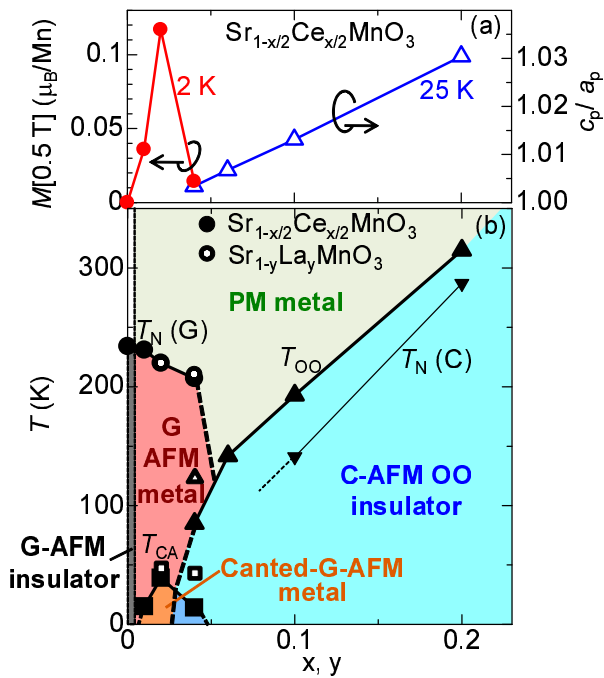}
\caption{\label{fig:phase}(Color online) (a) $M$ at 0.5 T at 2 K and tetragonality ($c_{\rm p}/a_{\rm p}$) at 25 K versus $x$ for Sr$_{1-x/2}$Ce$_{x/2}$MnO$_{3}$ ($0\!\le\!x\!\le\!0.2$), where $a_{\rm p}$ and $c_{\rm p}$ are the lattice constants defined in the perovskite subcell. (b) Electronic phase diagram as a function of $x$, based on the data for warming runs in Figs. \ref{fig:MTRT}. The transition temperatures of G-type [$T_{\rm N}$(G)], canted G-type ($T_{\rm CA}$) and C-type AFM [$T_{\rm N}$(C)] are indicated by closed circles, squares, and inverted triangles, respectively. The transition to orbital-ordered (OO) phase is represented by closed triangles ($T_{\rm OO}$). The data for Sr$_{1-y}$La$_{y}$MnO$_{3}$ ($y$=0.02, 0.04) are indicated by the corresponding open symbols.}
\end{figure}
%%%%%%%%%%%%%%%%%%%%%%%%%%%%%%%%%%%%%%%%%%%%%%%%%%%%%%%%%%%%%%%%%%%%%%%%%%%%%%%%%%%%%%%%%%%%%%%%%%%
%
%%%%%%%%%%%%%%%%%%%%%%%%%%%%%%%%%%%%%%%%%%%%%%%%%%%%%%%%%%%%%%%%%%%%%%%%%%%%%%%%%%%%%%%%%%%%%%%%%%%
\par
%%%%%%%%%%%%%%%%%%%%%%%%%%%%%%%%%%%%%%%%%%%%%%%%%%%%%%%%%%%%%%%%%%%%%%%%%%%%%%%%%%%%%%%%%%%%%%%%%%%
%
%%%%%%%%%%%%%%%%%%%%%%%%%%%%%%%%%%%%%%%%%%%%%%%%%%%%%%%%%%%%%%%%%%%%%%%%%%%%%%%%%%%%%%%%%%%%%%%%%%%
These results are summarized as an electronic phase diagram for Sr$_{1-x/2}$Ce$_{x/2}$MnO$_{3}$ ($0\!\le\!x\!\le\!0.2$) as a function of $x$ [Fig. \ref{fig:phase}(b)].
It reveals the critical phase competition between G-type AFM metal and C-type AFM OO insulator.
Both $T_{\rm N}$(G) and $T_{\rm OO}$ [or $T_{\rm N}$(C)] systematically decrease toward the bicritical point ($x_{\rm c}\!\sim\!0.05$), although the OO phase extends in the G-AFM phase below $\sim$100 K.
In the G-type AFM metal, the spin canting evolves with increasing $x$ up to 0.02, showing the increase in the spontaneous $M$ [Fig. \ref{fig:phase}(a)] and  $T_{\rm CA}$.
Further increase in $x$ toward the phase boundary to the C-type AFM OO, however, strongly reduces the canting.
%%%%%%%%%%%%%%%%%%%%%%%%%%%%%%%%%%%%%%%%%%%%%%%%%%%%%%%%%%%%%%%%%%%%%%%%%%%%%%%%%%%%%%%%%%%%%%%%%%%
%
%%%%%%%%%%%%%%%%%%%%%%%%%%%%%%%%%%%%%%%%%%%%%%%%%%%%%%%%%%%%%%%%%%%%%%%%%%%%%%%%%%%%%%%%%%%%%%%%%%%
\par
%%%%%%%%%%%%%%%%%%%%%%%%%%%%%%%%%%%%%%%%%%%%%%%%%%%%%%%%%%%%%%%%%%%%%%%%%%%%%%%%%%%%%%%%%%%%%%%%%%%
%
%%%%%%%%%%%%%%%%%%%%%%%%%%%%%%%%%%%%%%%%%%%%%%%%%%%%%%%%%%%%%%%%%%%%%%%%%%%%%%%%%%%%%%%%%%%%%%%%%%%
We first focus on the C-AFM OO insulating phase, which dominates the above phase diagram.
As shown in Fig. \ref{fig:x-ray}(a), the powder x-ray diffraction patterns at 25 K for $x\!\le\!0.02$ are nicely indexed with the cubic structure ($Pm\bar{3}m$), while those for $x\!\ge\!0.04$ show the peak splitting, reflecting the structural change into the tetragonal phase.
In fact, the profiles for $x\!\ge\!0.1$ well correspond to the space group $I4/mcm$ with the lattice constants $a\!\sim\!\sqrt{2}a_{\rm p}$, $c\!\sim\!2c_{\rm p}$ ($a_{\rm p}$, $c_{\rm p}$: lattice constants in the pseudo-cubic setting) \cite{Sundaresan2000EPJBa,Hervieu2000ChemMata,Chmaissem2003PRBa}, while the unit cell for $x\!\le\!0.06$ can be assigned with $a\!\sim\!a_{\rm p}$, $c\!\sim\!c_{\rm p}$.
As shown in Fig. \ref{fig:x-ray}(b), the tetragonality ($c_{\rm p}/a_{\rm p}$) rapidly increases just below $T_{\rm OO}$ and then saturates toward the lowest $T$.
Its value almost linearly increases with increasing $x$ up to $x$=0.2 [Fig. \ref{fig:phase}(a)].
%%%%%%%%%%%%%%%%%%%%%%%%%%%%%%%%%%%%%%%%%%%%%%%%%%%%%%%%%%%%%%%%%%%%%%%%%%%%%%%%%%%%%%%%%%%%%%%%%%%%%%%%
%
%%%%%%%%%%%%%%%%%%%%%%%%%%%%%%%%%%%%%%%%%%%%%%%%%%%%%%%%%%%%%%%%%%%%%%%%%%%%%%%%%%%%%%%%%%%%%%%%%%%
%%%%%%%%%%%%%%%%%%%%%%%%%%%%%%%%%%%%%%%%%%%%%%%%%%%%%%%%%%%%%%%%%%%%%%%%%%%%%%%%%%%%%%%%%%%%%%
%
%%%%%%%%%%%%%%%%%%%%%%%%%%%%%%%%%%%%%%%%%%%%%%%%%%%%%%%%%%%%%%%%%%%%%%%%%%%%%%%%%%%%%%%%%%%%%%%%%%%
\par
%%%%%%%%%%%%%%%%%%%%%%%%%%%%%%%%%%%%%%%%%%%%%%%%%%%%%%%%%%%%%%%%%%%%%%%%%%%%%%%%%%%%%%%%%%%%%%%%%%%
%
%%%%%%%%%%%%%%%%%%%%%%%%%%%%%%%%%%%%%%%%%%%%%%%%%%%%%%%%%%%%%%%%%%%%%%%%%%%%%%%%%%%%%%%%%%%%%%%%%%%
Noteworthy is that only $\sim$4\% electrons lead to the anisotropic OO ground state in otherwise the isotropic (cubic) compound. 
Compared to CaMnO$_{3}$, SrMnO$_{3}$ has much stronger OO instability against electron doping.
In Ca$_{1-x/2}$Ce$_{x/2}$MnO$_{3}$, the canted G-type AFM metallic phase prevails for $0.05\!\le\!x\!\le\!0.15$ in spite of the narrower bandwidth; more than 15\% electrons are necessary for the OO ground state \cite{Zeng2001PRBa,Caspi2004PRBa}.
Such a difference is considered to originate from the difference in the degree of $e_{g}$ orbital degeneracy as follows:
In cubic SrMnO$_{3}$ with totally degenerate orbitals, the Jahn-Teller effect may be dominant over the kinetic energy gain for doped electrons.
In CaMnO$_{3}$, on the other hand, the energy splitting of the $e_{g}$ orbitals due to the orthorhombic distortion makes the Jahn-Teller instability, in particular the collective distortion by the chainlike $3z^{2}\!-\!r^{2}$ polaron order, less effective; as a result the canted-AFM metal appears to be stabilized up to higher $x$.
The comparison of the two systems thus indicates that the degree of orbital degeneracy can play a vital role in the DE system via the Jahn-Teller effect, as also highlighted in the metallic phase ({\it vide infra}).
%%%%%%%%%%%%%%%%%%%%%%%%%%%%%%%%%%%%%%%%%%%%%%%%%%%%%%%%%%%%%%%%%%%%%%%%%%%%%%%%%%%%%%%%%%%%%%%%%%%
%
%%%%%%%%%%%%%%%%%%%%%%%%%%%%%%%%%%%%%%%%%%%%%%%%%%%%%%%%%%%%%%%%%%%%%%%%%%%%%%%%%%%%%%%%%%%%%%%%%%%
\par
%%%%%%%%%%%%%%%%%%%%%%%%%%%%%%%%%%%%%%%%%%%%%%%%%%%%%%%%%%%%%%%%%%%%%%%%%%%%%%%%%%%%%%%%%%%%%%%%%%%
%
%%%%%%%%%%%%%%%%%%%%%%%%%%%%%%%%%%%%%%%%%%%%%%%%%%%%%%%%%%%%%%%%%%%%%%%%%%%%%%%%%%%%%%%%%%%%%%%%%%%
We next scrutinize the G-type AFM metallic phase, which shows up only at a low electron concentration of 1-2\%.
[$T_{\rm N}$(G) is hereafter simplified into $T_{\rm N}$.]
Figure \ref{fig:polaron}(a) magnifies the metallic $\rho$-$T$ curves for Sr$_{1-x/2}$Ce$_{x/2}$MnO$_{3}$ as well as Sr$_{1-y}$La$_{y}$MnO$_{3}$.
In both systems, qualitatively the same behavior has been observed for the same electron density, irrespective of the doping species, i.e., the magnitude of disorder effects arising from the A-site solid solution \cite{Tokura2006review}.
For $x$ or $y$=0.01-0.02, the $\rho$-$T$ curve shows a distinct change in its slope at $T_{\rm N}$.
Below $T_{\rm N}$, the long-range AFM ordering of $t_{2g}$ spins reduces the scattering of conduction electrons and hence the resistivity with decreasing $T$.
For $x$ or $y$=0.04, the similar metallic behavior was observed around $T_{\rm N}$, which is followed by a steep increase in $\rho$ due to the OO transition at lower $T$.
%%%%%%%%%%%%%%%%%%%%%%%%%%%%%%%%%%%%%%%%%%%%%%%%%%%%%%%%%%%%%%%%%%%%%%%%%%%%%%%%%%%%%%%%%%%%%%%%%%%
%
%%%%%%%%%%%%%%%%%%%%%%%%%%%%%%%%%%%%%%%%%%%%%%%%%%%%%%%%%%%%%%%%%%%%%%%%%%%%%%%%%%%%%%%%%%%%%%%%%%%
%%%%%%%%%%%%%%%%%%%%%%%%%%%%%%%%%%%%%%%%%%%%%%%%%%%%%%%%%%%%%%%%%%%%%%%%%%%%%%%%%%%%%%%%%%%%%%%%%%%
%
%%%%%%%%%%%%%%%%%%%%%%%%%%%%%%%%%%%%%%%%%%%%%%%%%%%%%%%%%%%%%%%%%%%%%%%%%%%%%%%%%%%%%%%%%%%%%%%%%%%
\par
%%%%%%%%%%%%%%%%%%%%%%%%%%%%%%%%%%%%%%%%%%%%%%%%%%%%%%%%%%%%%%%%%%%%%%%%%%%%%%%%%%%%%%%%%%%%%%%%%%%
%
%%%%%%%%%%%%%%%%%%%%%%%%%%%%%%%%%%%%%%%%%%%%%%%%%%%%%%%%%%%%%%%%%%%%%%%%%%%%%%%%%%%%%%%%%%%%%%%%%%%
Figure \ref{fig:polaron}(b) shows the $T$ profiles of $C$ below 10 K for $x\!=\!0.01$ and 0.02, plotted as $C/T$ versus $T^{2}$.
The data were well fit with a relation $C/T\!=\!\gamma\!+\!\beta T^{2}$ (solid lines), resulting in the electronic specific heat coefficients $\gamma\!\sim\!5.4$ ($x$=0.01) and $\sim$5.7 ($x$=0.02) mJ/K$^{2}$/mol.
These values give heavy effective masses, $m^{*}\!\sim\!11m_{0}$ ($x$=0.01) and $\sim\!9.6m_{0}$ ($x$=0.02), provided that the $e_{g}$ bands are doubly degenerate parabolic ones ($m_{0}$: free electron mass).
The corresponding $Q$ (Seebeck coefficient)-$T$ curves show steep gradient at low $T$ [Fig. \ref{fig:polaron}(c)].
This behavior is well reproduced by the Boltzmann transport theory (solid lines), using the above $m^{*}$ values and the constant relaxation time \cite{Okuda2001PRBa}.
Around 140 K, however, the experimental $|Q|$ values begin to decrease with increasing $T$, showing the deviation from the theoretical lines due to the thermal fluctuation evolving toward $T_{\rm N}$.
Well above $T_{\rm N}$, again, they gradually increase with increasing $T$, which are nicely fit with the reduced $m^{*}\!\sim\!3.1m_{0}$ ($x$=0.01) and $\sim\!2.6m_{0}$ ($x$=0.02), as shown by dashed lines.
From the band calculation based on the spin-unpolarized local density approximation \cite{WIEN2K}, the $e_{g}$ band mass $m_{b}$ is estimated to be $\sim$$0.65m_{0}$ \cite{bandmass}, which leads to $m^{*}/\langle m_{b}\rangle$$\sim$3.0 ($x$=0.01) and 2.6 ($x$=0.02) for $T\!\gg\!T_{\rm N}$.
Note here that $\langle m_{b} \rangle\!\!\equiv\![(1/\pi)\int^{\pi}_{0}\cos{(\theta/2)}d\theta]^{-1}m_{b}\!=\!(\pi/2)m_{b}$ means the thermal average of the angle ($\theta$) between the fluctuating $t_{2g}$ spins in the PM phase.
The origin of such significant mass renormalization of dilute carriers can be assigned to the strong electron-phonon interaction, as suggested also in CaMnO$_{3}$ \cite{Cohn2002PRBa}. 
This coupling, presumably with the Jahn-Teller phonons, would be further promoted in the present orbital-degenerate cubic SrMnO$_{3}$ system.
%%%%%%%%%%%%%%%%%%%%%%%%%%%%%%%%%%%%%%%%%%%%%%%%%%%%%%%%%%%%%%%%%%%%%%%%%%%%%%%%%%%%%%%%%%%%%%%%%%%
%
%%%%%%%%%%%%%%%%%%%%%%%%%%%%%%%%%%%%%%%%%%%%%%%%%%%%%%%%%%%%%%%%%%%%%%%%%%%%%%%%%%%%%%%%%%%%%%%%%%%
\par
%%%%%%%%%%%%%%%%%%%%%%%%%%%%%%%%%%%%%%%%%%%%%%%%%%%%%%%%%%%%%%%%%%%%%%%%%%%%%%%%%%%%%%%%%%%%%%%%%%%
%
%%%%%%%%%%%%%%%%%%%%%%%%%%%%%%%%%%%%%%%%%%%%%%%%%%%%%%%%%%%%%%%%%%%%%%%%%%%%%%%%%%%%%%%%%%%%%%%%%%%
The enhancement in $m^{*}$ below $T_{\rm N}$ can be explained by the reduction of the one-electron bandwidth $W$ in the G-type AFM phase, where only the second (or higher-order) nearest-neighbor (NN) hopping is virtually allowed due to the large Hund's rule coupling energy.
When the value of $W$ ($T\!\gg\!T_{\rm N}$) is reduced to $W^{\prime}$ ($T\!\ll\!T_{\rm N}$) [$r\!=\!W^{\prime}/W (\le1)$], the expression of mass renormalization $f(\lambda)$ changes as follows ($\lambda$: electron-phonon coupling constant): $m^{*}/\langle m_{b}\rangle\!=\!f(\lambda)$ ($T\!\gg\!T_{\rm N}$) $\rightarrow$ $m^{*\prime}/m_{b}^{\prime}\!=\!f(\lambda^{\prime})$ ($T\!\ll\!T_{\rm N}$), where the prime mark denotes the value for $T\!\ll\!T_{\rm N}$, and $\lambda^{\prime}\!=\!\lambda/\sqrt{r}$ \cite{lambda}, $m_{b}^{\prime}\!=\!m_{b}/r$.
Note here that no thermal average is taken at $T\!\ll\!T_{\rm N}$.
Based on $f(\lambda)$ calculated in the Fr\"{o}hlich model \cite{Mishchenko2000PRBa}, we have obtained $(\lambda,r)$=$(0.78,0.50)$ for $x$=0.01 and $(0.69,0.45)$ for $x$=0.02, to reproduce the experimental $m^{*}$ (and $m^{*\prime}$) values.
On the other hand, the $r$ value can be directly estimated from $r\!=\!2t_{2}/\langle t_{1}+2t_{2}\rangle\!=\!(\pi/2)[2t_{2}/(t_{1}+2t_{2})]$, where $t_{1}$ and $t_{2}$ are the NN and 2nd-NN hopping parameters, respectively, and $2t_{2}$ comes from twice larger coordination number for the second-NN sites than for the NN.
For $t_{1}$=0.5-0.75 (eV) and $t_{2}$=0.2-0.3 (eV), as employed in \cite{Meskine2004PRLa}, we have $r$=0.55-0.86; the lower limit case roughly coincides with the above result deduced from the $m^{*}$ change.
%%%%%%%%%%%%%%%%%%%%%%%%%%%%%%%%%%%%%%%%%%%%%%%%%%%%%%%%%%%%%%%%%%%%%%%%%%%%%%%%%%%%%%%%%%%%%%%%%%%
%
%%%%%%%%%%%%%%%%%%%%%%%%%%%%%%%%%%%%%%%%%%%%%%%%%%%%%%%%%%%%%%%%%%%%%%%%%%%%%%%%%%%%%%%%%%%%%%%%%%%
\par
%%%%%%%%%%%%%%%%%%%%%%%%%%%%%%%%%%%%%%%%%%%%%%%%%%%%%%%%%%%%%%%%%%%%%%%%%%%%%%%%%%%%%%%%%%%%%%%%%%%
%
%%%%%%%%%%%%%%%%%%%%%%%%%%%%%%%%%%%%%%%%%%%%%%%%%%%%%%%%%%%%%%%%%%%%%%%%%%%%%%%%%%%%%%%%%%%%%%%%%%%
In spite of the strong electron-phonon coupling ($\lambda\!\sim\!1$), the dilute gas of Jahn-Teller polarons in SrMnO$_{3}$ shows the metallic conduction, free from the impurity-assisted self-trapping process \cite{Shinozuka1979JPSJa}.
This indicates the effective screening of the ionized impurities, which is indeed supported by the high dielectric constant $\varepsilon_{r}$ observed in SrMnO$_{3}$ ($\varepsilon_{r}\!\sim\!110$ at 1 MHz at 5 K).
Recent theories \cite{Bhattacharjee2009PRLa, Rondinelli2009PRBa} have further predicted that the ferroelectric instability develops in SrMnO$_{3}$, as in the case of quantum paraelectric SrTiO$_{3}$.
Thus, the emergence of the isotropic metal originates from the complex interplay among the magnetic exchange interaction, electron-phonon coupling and high dielectric response.
%%%%%%%%%%%%%%%%%%%%%%%%%%%%%%%%%%%%%%%%%%%%%%%%%%%%%%%%%%%%%%%%%%%%%%%%%%%%%%%%%%%%%%%%%%%%%%%%%%%
%
%%%%%%%%%%%%%%%%%%%%%%%%%%%%%%%%%%%%%%%%%%%%%%%%%%%%%%%%%%%%%%%%%%%%%%%%%%%%%%%%%%%%%%%%%%%%%%%%%%%
\begin{acknowledgments}
%%%%%%%%%%%%%%%%%%%%%%%%%%%%%%%%%%%%%%%%%%%%%%%%%%%%%%%%%%%%%%%%%%%%%%%%%%%%%%%%%%%%%%%%%%%%%%%%%%%
%
%%%%%%%%%%%%%%%%%%%%%%%%%%%%%%%%%%%%%%%%%%%%%%%%%%%%%%%%%%%%%%%%%%%%%%%%%%%%%%%%%%%%%%%%%%%%%%%%%%%
We appreciate J. Fujioka and A. Mishchenko for fruitful discussions.
We also thank M. Uchida for his help in the high-$T$ $Q$ measurement and Y. Shiomi and N. Kanazawa for the band calculation.
This study was partly supported by Special Postdoctoral Program in RIKEN, KAKENHI 22740244, Photon Factory Program (No. 2009S2-008) and Funding Program for World-Leading Innovative R\&D on Science and Technology (FIRST) on ``Strong-correlation quantum science''.
%%%%%%%%%%%%%%%%%%%%%%%%%%%%%%%%%%%%%%%%%%%%%%%%%%%%%%%%%%%%%%%%%%%%%%%%%%%%%%%%%%%%%%%%%%%%%%%%%%%
%
%%%%%%%%%%%%%%%%%%%%%%%%%%%%%%%%%%%%%%%%%%%%%%%%%%%%%%%%%%%%%%%%%%%%%%%%%%%%%%%%%%%%%%%%%%%%%%%%%%%
\end{acknowledgments}
%%%%%%%%%%%%%%%%%%%%%%%%%%%%%%%%%%%%%%%%%%%%%%%%%%%%%%%%%%%%%%%%%%%%%%%%%%%%%%%%%%%%%%%%%%%%%%%%%%%
%
%%%%%%%%%%%%%%%%%%%%%%%%%%%%%%%%%%%%%%%%%%%%%%%%%%%%%%%%%%%%%%%%%%%%%%%%%%%%%%%%%%%%%%%%%%%%%%%%%%%

%%%%%%%%%%%%%%%%%%%%%%%%%%%%%%%%%%%%%%%%%%%%%%%%%%%%%%%%%%%%%%%%%%%%%%%%%%%%%%%%%%%%%%%%%%%%%%%%%%%
%
%%%%%%%%%%%%%%%%%%%%%%%%%%%%%%%%%%%%%%%%%%%%%%%%%%%%%%%%%%%%%%%%%%%%%%%%%%%%%%%%%%%%%%%%%%%%%%%%%%%

\begin{thebibliography}{}
\bibitem{phaseseparation} Macroscopic phase separation is also unlikely, since the similar metallic conduction was observed for several specimens. No impurity phase was detected in the synchrotron x-ray diffraction profiles (Figs. \ref{fig:x-ray}).
\end{thebibliography}
\end{document}